\documentstyle[pre,aps]{revtex}
\input epsf
\tighten
\begin{document}
\draft
\twocolumn[\hsize\textwidth\columnwidth\hsize\csname @twocolumnfalse\endcsname
\title{Numerical study of discrete models in the
class of the nonlinear molecular beam epitaxy equation}
\author{F. D. A. Aar\~ao Reis}
\address{Instituto de F\'\i sica, Universidade Federal Fluminense,\\
Avenida Litor\^anea s/n, 24210-340 Niter\'oi RJ, Brazil}
\date{\today}
\maketitle
\begin{abstract}
We study numerically some discrete growth models belonging to the class of the
nonlinear molecular beam epitaxy equation, or Villain-Lai-Das Sarma
(VLDS) equation. The conserved restricted solid-on-solid model (CRSOS) with
maximum heights differences ${\Delta H}_{max}=1$ and ${\Delta H}_{max}=2$ was
analyzed in substrate dimensions $d=1$ and $d=2$. The Das Sarma and
Tamborenea (DT) model and a competitive model involving random deposition and
CRSOS deposition were studied in $d=1$. For the CRSOS model with ${\Delta
H}_{max}=1$ we obtain the more accurate estimates of scaling exponents in
$d=1$: roughness exponent $\alpha =0.94\pm 0.02$ and dynamical
exponent $z=2.88\pm 0.04$. These estimates are significantly below the values
of one-loop renormalization for the VLDS theory, which confirms Janssen's
proposal of the existence of higher order corrections. The roughness exponent
in $d=2$ is very near the one-loop result $\alpha=2/3$, in agreement with
previous works. The moments $W_n$ of orders $n=2,3,4$ of the heights
distribution were calculated for all models and the skewness $S \equiv {W_3}
/{{W_2}^{3/2}}$ and the kurtosis $Q \equiv {W_4} /{{W_2}^{2}} -3$ were
estimated. At the steady states, the CRSOS models and the competitive model
have nearly the same values of $S$ and $Q$ in $d=1$, which suggests that these
amplitude ratios are universal in the VLDS class. The estimates for the DT
model are different, possibly due to their typically long crossover to
asymptotic values. Results for the CRSOS models in $d=2$ also suggest that
those quantities are universal.
\end{abstract}


\narrowtext
\vskip2pc]

\section{Introduction}

Surface and interface growth processes are subjects of great interest for the
perspective of applications to thin films and multilayers growth and, from the
theoretical point of view, for their important role in Non-Equilibrium Statistical
Mechanics~\cite{barabasi,krug}. Frequently those processes are described by
discrete models which represent the basic growth mechanisms by simple
stochastic rules, such as aggregation and diffusion, and neglect details of
the microscopic interactions. On the other hand, continuous theories are
successful at representing those processes in the hydrodynamic limit. They
predict the scaling exponents of many discrete models, which are consequently
grouped in a small number of universality classes.

Growth by molecular beam epitaxy ($MBE$), which is one of the most important
techniques to produce high quality films with smooth surfaces, motivated the
proposal of many discrete and continuous models. The dynamics
during MBE deposition is dominated by diffusion processes, which led to the
proposal of an important theoretical model, the Villain-Lai-Das Sarma (VLDS)
growth equation~\cite{villain,laidassarma}
\begin{equation}
{{\partial h}\over{\partial t}} = \nu_4{\nabla}^4 h +
\lambda_{4} {\nabla}^2 {\left( \nabla h\right) }^2 + \eta (\vec{x},t) ,
\label{vlds}
\end{equation}
where $h(\vec{x},t)$ is the height at position $\vec{x}$ and time $t$ in a
$d$-dimensional substrate, $\nu_4$ and $\lambda_{4}$ are constants and $\eta$
is a Gaussian (nonconservative) noise. Eq. (\ref{vlds}) is also frequently
called nonlinear molecular beam epitaxy equation or conserved
Kardar-Parisi-Zhang equation~\cite{barabasi,kpz},

The most important geometrical quantity to characterize the surface of the
deposit grown by such processes is the interface width. It is defined as the
root mean square fluctuation of the average height
\begin{equation}
\xi \equiv {\left[ \left< { \left( h - \overline{h}\right) }^2
\right> \right] }^{1/2} .
\label{defw}
\end{equation}
For short times, it scales as
\begin{equation}
\xi\sim t^{\beta} ,
\label{defbeta}
\end{equation}
where $\beta$ is called the growth exponent.
For long times, in the steady state, the interface width saturates at
\begin{equation}
\xi_{sat}\sim L^{\alpha} ,
\label{defalpha}
\end{equation}
where $\alpha$ is called the roughness exponent.
The crossover time from the growth regime to the steady state scales with $L$
with the dynamical exponent
\begin{equation}
z=\alpha /\beta .
\label{defz}
\end{equation}

For the VLDS theory, a one-loop dynamical renormalization-group (DRG)
calculation~\cite{villain,laidassarma} led to $\alpha=(4-d)/3$, $z=(8+d)/3$
and $\beta=(4-d)/(8+d)$ below the upper critical dimension $d_c=4$. See also
the recent work of Katzav~\cite{katzav}, based on a self-consistent expansion
approach, which also obtains these estimates. Some authors assumed
the one-loop values to be exact in all orders, but Janssen~\cite{janssen}
recently claimed that this conclusion was derived from
an ill-defined transformation and, consequently, there would be higher order
corrections. From a two-loop calculation, he obtained small negative
corrections to $\alpha$ and $z$ in all dimensions~\cite{janssen}. Numerical
studies of some discrete models which belong to the VLDS class in the
continuum limit (large lattices, long times) were not able to solve this
controversy. In $d=1$, numerical work on a conserved restricted solid-on-solid
model (to be defined below) systematically suggest
$\alpha<1$~\cite{crsos1,kimkim97}, but the error bars are large and,
consequently, the authors still suggest the validity of the one-loop result.
In $d=2$ and higher dimensions~\cite{yook}, numerical results indicated that
possible corrections to the one-loop result were smaller than the two-loops
estimates of Janssen~\cite{janssen}.

Another important question is motivated by recent results
on discrete models belonging to the Kardar-Parisi-Zhang (KPZ) class in $d=2$.
The KPZ growth equation includes second order linear and nonlinear terms
which are more relevant than those in the VLDS equation (Eq. \ref{vlds})
in the hydrodynamic limit~\cite{kpz,barabasi}. Works on discrete KPZ models
showed that the steady state values of the moments of the height
distribution,
\begin{equation}
W_n \equiv {\left< \overline{ {\left( h-\overline{h}\right) }^n } \right> } ,
\label{defmoments}
\end{equation}
obey power-counting, i. e. they scale as
\begin{equation}
W_n \sim L^{n\alpha} 
\label{powercounting}
\end{equation}
(note that $W_2=\xi^2$).
Moreover, estimates of the skewness
\begin{equation}
S \equiv {{W_3}\over{{W_2}^{3/2}}}
\label{defskew}
\end{equation}
and of the kurtosis
\begin{equation}
Q \equiv {{W_4}\over{{W_2}^{2}}}-3
\label{defkurt}
\end{equation}
of the KPZ models indicated that the amplitude ratios of the moments $W_n$
(such as $S$ and $Q$) are universal~\cite{chin,marinari,kpz2d}. It seems that
no previous work has considered these questions in models belonging to the
VLDS class, possibly due to the large times involved in their simulations (the
dynamical exponent is nearly the double of the KPZ value). Besides the
theoretical relevance of those questions, additional motivation for
their analysis is the fact that the amplitude ratios can be measured with
much higher accuracy than the scaling exponents and may eventually help one to
infer the universality class of an experimental growth process.

There is a small number of discrete models belonging to the VLDS class in the
continuum limit. The discrete model proposed by Das Sarma and Tamborenea ($DT$
model)~\cite{dt} is an example of a MBE-motivated model which falls in that
class in $d=1$, although  there is evidence that its class in
$d=2$ is different~\cite{toroczkai,chamereis}. On the other hand, the
so-called conserved restricted-solid-on-solid (CRSOS) models, first proposed
by Kim et al~\cite{crsos1}, is expected to belong to the VLDS class in all
dimensions. This was already proved analytically in
$d=1$~\cite{huang,park1,park2}. In the CRSOS models, the difference of the
heights of neighboring columns are always smaller than a certain value
${\Delta H}_{max}$, similarly to the RSOS model of Kim and
Kosterlitz~\cite{kk,kkala}. However, in the Kim-Kosterlitz model, if
the aggregation at the column of incidence does not satisfy that condition,
then the aggregation attempt is rejected (consequently, the model is in the KPZ
class). On the other hand, in the CRSOS model, the incident particle migrates
to the nearest column at which the height difference constraint is satisfied
after aggregation. Thus, all deposition attempts are successful in the CRSOS
model.

Here, we will study numerically a modified version of the CRSOS model in $d=1$
and $d=2$, with two different values of ${\Delta H}_{max}$, the DT model in
$d=1$, simulated with noise-reduction methods, and a competitive model
involving CRSOS and random deposition in $d=1$. All these models
belong to the VLDS class. We will perform systematic extrapolations of
effective (roughness and dynamical) exponents for the CRSOS model in $d=1$ and
$d=2$. The asymptotic exponents in $d=1$ are clearly different
from the one-loop DRG values and the sign of the deviations are in qualitative
agreement with Janssen's results~\cite{janssen}. In $d=2$, possible
corrections in the exponent $\alpha$ are smaller than the two-loop corrections
calculated in that work, confirming other authors' conclusions. It will also
be shown that the moments of the heights distribution obey power-counting (Eq.
\ref{powercounting}) in $d=1$ and $d=2$, similarly to KPZ, and that the
skewness and the kurtosis for different versions of the CRSOS model (different
${\Delta H}_{max}$) and for the competitive model have nearly the same values.
These estimates differ from those of the DT model in $d=1$, but universality
of amplitude ratios in the VLDS class cannot be discarded due to the typical
long crossovers of the DT model.

The rest of this paper is organized as follows. In Sec. II we present the
stochastic rules of the CRSOS and DT models and give information on the
simulation procedure. In Sec. III, we calculate the scaling exponents of the
VLDS class in one-dimensional substrates. In Sec. IV, we calculate the scaling
exponents in two-dimensional substrates. In Sec. V, we compare the
asymptotic amplitude ratios of all models in $d=1$ and $d=2$. In Sec. VI we
summarize our results and present our conclusions.

\section{Models and simulation procedure}

The rules for choosing the aggregation point in our version of the CRSOS model
are slightly different from the original ones. The present version
was introduced in Ref. \protect\cite{cnpre} as a model for amorphous
carbon-nitrogen films growth, but only small lattices were analyzed there and,
consequently, reliable estimates of scaling exponents were not obtained.

At any time, all pairs of neighboring columns are restricted to obey the
condition $\Delta h\leq {\Delta H}_{max}$, where $\Delta h$ is the difference
in the columns' heights and ${\Delta H}_{max}$ is fixed. The deposition
attempt begins with the random choice of one substrate column $i$. If the
above condition is satisfied after aggregation of a new particle at the top of
column $i$, then the aggregation takes place at that position. Otherwise, a
nearest neighbor column is randomly chosen (independently of its height) and
the same test is performed. This process is continued until a column is chosen
in which the new particle can be permanently deposited. Here, the cases
${\Delta H}_{max}=1$ and ${\Delta H}_{max}=2$ will be analyzed.

In the original version of the CRSOS model~\cite{crsos1}, the aggregation
takes place at the nearest column in which the condition on heights
differences is satisfied, but in our version the incident particle performs a
random walk along the substrate direction(s) while it searches for the
aggregation point. The original model was proved to belong to the VLDS class
in $d=1$ by different methods~\cite{huang,park1,park2} and the coefficients of
the VLDS equation were explicitly calculated for ${\Delta
H}_{max}=1$~\cite{park1,park2}. Since our version does not change any symmetry
of the original CRSOS model, it is also expected to be in that class. Notice,
for instance, that there is no upward or downward current in our model due to
the mechanism of random walks for choosing the aggregation position (the
random steps do not depend on the relative heights of the columns). It
implies that the coefficient of the second order height derivative of the
growth equation (not shown in Eq. \ref{vlds}) is exactly zero, the VLDS
equation being the most plausible continuum description - see e. g. the
discussion in Ref. \protect\cite{hagston}.

We will also study the DT model in $d=1$. In this model, the
incident particle sticks at the top of the randomly chosen column $i$ if it
has one or two lateral neighbors at that position (a kink site or a valley,
respectively). Otherwise, the neighboring columns (at the right and the left
sides in $d=1$) are consulted. If the top position of only one of these
columns is a kink site or a valley, then the incident particle aggregates at
that point. If no neighboring column satisfies that condition, then the
particle sticks at the top of column $i$. Finally, if both neighboring columns
satisfy that condition, then one of them is randomly chosen.

In our simulations of the DT model, we used the noise reduction
technique adopted in Ref. \protect\cite{punyindu}. The noise reduction factor
$m$ is the number of attempts at a site for an actual aggregation
process to occur~\cite{wolf,kertesz}. Here, the value $m=10$ will
be considered because it provided accurate estimates of scaling exponents in
Ref. \protect\cite{punyindu} from simulations in relatively small
systems. On the other hand, the data for the original DT model
present huge finite-size corrections (see e. g. Ref. \protect\cite{brunoc}).

In order to improve our discussion on the universality of amplitude ratios 
(Sec. V), we also simulated a competitive model in which the aggregation of
the incident particle may follow two different rules: with probability $p$,
the particle aggregates at the top of the column of incidence, such as in the
random deposition (RD) model~\cite{barabasi}; otherwise (probability $1-p$), it
diffuses until finding a column $i$ in which the condition $h_i-h_j\leq{\Delta
H}_{max}$ is satisfied for all nearest neighbors $j$ after aggregation. Thus,
the latter aggregation mechanism works for preserving the columns heights'
constraint of the CRSOS model. Extending previous conclusions on other
competitive models~\cite{albano1,albano2}, it is expected that this model is
described asymptotically by the VLDS equation, similarly to the pure CRSOS
model, but the coefficients $\nu_4$ and $\lambda_4$ of the corresponding
continuous equation (Eq. \ref{vlds}) are expected to depend on $p$. In this
paper, we will simulate the model with $p=0.25$ ($p=0$ is the pure CRSOS
model).

The above models were simulated in $d=1$ in lattices of lengths ranging from
$L=16$ to $L=1024$ for the CRSOS model with ${\Delta H}_{max}=1$ and ${\Delta
H}_{max}=2$, from $L=16$ to $L=256$ for the DT model and from $L=16$ to
$L=512$ for the competitive model. For the CRSOS models, the number of
realizations up to the steady state was typically ${10}^4$ for the smallest
lattices and nearly $500$ for the largest lattices. The same applies to the
DT model, but notice that the largest length in that case was just $L=256$. In
$d=2$, the CRSOS model with ${\Delta H}_{max}=1$ was simulated in lattices of
lengths ranging from $L=16$ to $L=256$, and with ${\Delta H}_{max}=2$ only
until $L=128$. Whenever the number of realizations up to the steady state was
smaller than ${10}^4$, a larger number of realizations covering the growth and
the crossover regions was generated. This allowed the calculation of crossover
times (see below) with good accuracy in $d=1$.

The calculation of the moments of the height distribution at the steady
states, $W_n$ (Eq. \ref{defmoments}), followed the same lines described in
Ref. \protect\cite{kpz2d}. In order to estimate dynamical exponents, we used a
recently proposed method to calculate a characteristic time $\tau_0$ which is
proportional to the time of relaxation to the steady state~\cite{tau}. For
fixed $L$, after calculating the saturation width $\xi_{sat}(L)$, $\tau_0$ is
defined through
\begin{equation}
\xi{\left( L,\tau_0\right)} = k\xi_{sat}{\left( L\right)} ,
\label{deftau0}
\end{equation}
with a constant $k\lesssim 1$. From the Family-Vicsek relation~\cite{famvic},
it is expected that~\cite{tau}
\begin{equation}
\tau_0\sim L^z .
\label{scalingtau0}
\end{equation}

Here, we estimated $\tau_0$ with $k$ ranging from $k=0.4$ to $k=0.7$. Since
the exponent $z$ is large, the characteristic times $\tau_0$ increase very
fast with $L$. Consequently, for large $k$, the accuracy of $\tau_0$ is low
in large lattices. On the other hand, for small $k$, the times $\tau_0$ in
small lattices are also very small (near $\tau_0=1$) and, consequently,
there are effects of the initial flat substrate. This is the reason why
we chose a restricted range of $k$ to analyze our data. 

\section{Scaling exponents in one-dimensional substrates}

In order to estimate the roughness exponent from the interface width
$\xi$, the first step is to calculate the effective exponents
\begin{equation}
\alpha_{\left( L,i\right)} \equiv { \ln \left[\xi_{sat}\left( L\right)
/\xi_{sat}\left( L/i\right)\right] \over \ln{i} } 
\label{defalphaL}
\end{equation}
for fixed $i$. It is expected that $\alpha_{\left( L,i\right)}\to \alpha$ for
any choice of $i$.

\begin{figure}
\epsfxsize=8,5cm
\begin{center}
\leavevmode
\epsffile{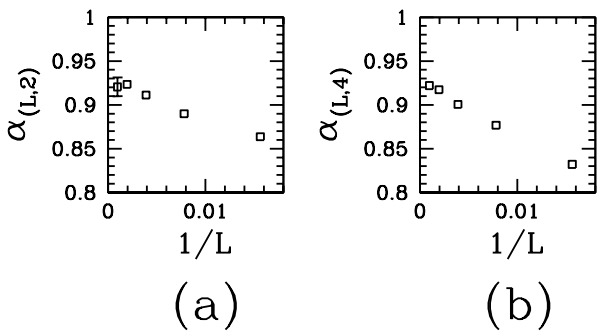}
\caption{Effective roughness exponents (a) $\alpha_{\left( L,2\right) }$ and
(b) $\alpha_{\left( L,4\right) }$ versus inverse lattice length for the
$1+1$-dimensional CRSOS model with ${\Delta H}_{max}=1$. Error bars are shown
only when they are larger than the size of the data points.}
\label{fig1}                        
\end{center}
\end{figure}     

In Figs. 1a and 1b we show $\alpha_{\left( L,2\right)}$ and $\alpha_{\left(
L,4\right)}$ versus $1/L$, respectively, for the CRSOS model with
${\Delta H}_{max}=1$. The evolution of the data suggests that
$\alpha_{\left( L,i\right)}$ converges to $0.91\leq \alpha\leq 0.94$,
accounting for the error bars and reasonable finite-size corrections.

The type of plot in Figs. 1a and 1b is suitable to fit the data to the
scaling form \begin{equation}
\alpha_{\left( L,i\right)} \approx \alpha + A L^{-\Delta} ,
\label{correctionalphaL}
\end{equation}
with $A$ constant, if the correct variable $L^{-\Delta}$ is used in the
abscissa ($\Delta =1$ was tested in Figs. 1a and 1b). In its
turn, Eq. (\ref{correctionalphaL}) is a consequence of a scaling relation
$\xi_{sat}\approx L^{\alpha} (a_0+a_1 L^{-\Delta})$, with $a_0$ and $a_1$
constants, which includes a sub-dominant term in addition to the dominant one
in Eq. (\ref{defalpha}). However, no variable
of the form $L^{-\Delta}$ provided a reasonable linear fit in the range of
lattice size analyzed there. Thus, $\Delta=1$ was used in Figs. 1a and 1b just
to illustrate the $L$-dependence of the effective exponents. On the other hand,
estimating the asymptotic $\alpha$ is possible because there is no evidence
of an upward curvature of those plots for large $L$.

\begin{figure}
\epsfxsize=8,5cm
\begin{center}
\leavevmode
\epsffile{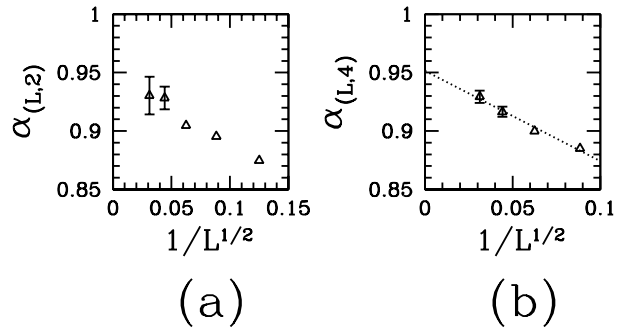}
\caption{Effective roughness exponents (a) $\alpha_{\left( L,2\right) }$ and (b)
$\alpha_{\left( L,4\right) }$ versus $1/L^{1/2}$ for the $1+1$-dimensional CRSOS
model with ${\Delta H}_{max}=2$. Error bars are shown only when they
are larger than the size of the data points.}
\label{fig2}                        
\end{center}
\end{figure}     

The data for the CRSOS model with ${\Delta H}_{max}=2$ was analyzed along the
same lines. In Figs. 2a and 2b we show $\alpha_{\left( L,2\right)}$ and $\alpha_{\left(
L,4\right)}$ versus $1/L^{1/2}$, respectively. The variable in the abscissa of
Figs. 2a and 2b was chosen to provide a good linear fit of the $\alpha_{\left(
L,4\right)}$ data - see dotted line in Fig. 2b. These results suggest
stronger finite-size corrections for $\alpha_{\left( L,i\right)}$ when
compared to the model with ${\Delta H}_{max}=1$. The corresponding
asymptotic estimates are in the range $0.92\leq \alpha \leq 0.97$, also
accounting for the error bars. However, since these error bars are larger than
those for ${\Delta H}_{max}=1$, it is possible that the true asymptotic
regime was not attained yet and that the true leading corrections are
different. Anyway, those results still suggest that $\alpha <1$ in the
$L\to\infty$ limit.

Alternatively, we will analyze our data assuming the
presence of a constant term as the sub-leading correction to the scaling of
${\xi_{sat}}^2$:
\begin{equation}
{\xi_{sat}}^2 = \xi_I^2 + A L^{2\alpha} .
\label{intrinsic}
\end{equation}
(since $\alpha\sim 1$, it corresponds asymptotically to $\Delta \sim 2$ in Eq.
\ref{correctionalphaL}). $\xi_I$ is called intrinsic width and is frequently
associated to large local slopes in discrete KPZ
models~\cite{wolf,kertesz,kpz2d}.  Effective exponents $\alpha_L^{\left(
I\right)}$ which cancel the contribution of $\xi_I^2$ may be defined as
\begin{equation}
\alpha_L^{\left( I\right)} \equiv {1\over 2}
{ \ln{ \left[ \xi_{sat}^2\left( 2L\right) - \xi_{sat}^2\left( L\right)
\right] /\left[ \xi_{sat}^2\left( L\right) - \xi_{sat}^2\left( L/2\right)
\right] } \over\ln{2} } .
\label{defalphai}
\end{equation}

\begin{figure}
\epsfxsize=8,5cm
\begin{center}
\leavevmode
\epsffile{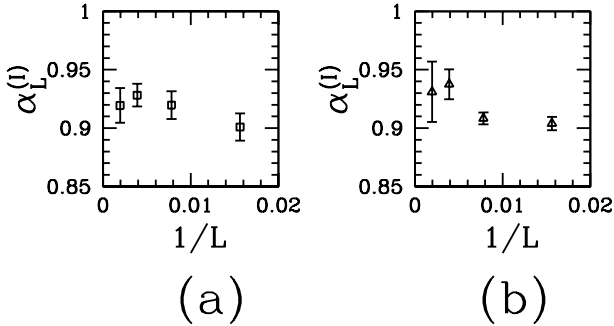}
\caption{Effective roughness exponents $\alpha_{L}^{(I)}$ (accounting for the
intrinsic width) versus $1/L$ for $1+1$-dimensional
CRSOS models with (a) ${\Delta H}_{max}=1$ and (b) ${\Delta H}_{max}=2$.}
\label{fig3}                        
\end{center}
\end{figure}     

In Figs. 3a and 3b we show $\alpha_L^{\left( I\right)}$ versus $1/L$ for the
CRSOS model with ${\Delta H}_{max}=1$ and ${\Delta H}_{max}=2$, respectively.
Here, the variable $1/L$ in the abscissa was also not chosen to perform data
extrapolation. The effective exponents vary within narrow ranges
($0.89$ to $0.94$ for ${\Delta H}_{max}=1$, $0.90$ to $0.96$ for ${\Delta
H}_{max}=2$), even including their error bars. Consequently, any variable in
the form $L^{-\Delta}$ ($0.5\leq \Delta\leq 2$) leads to nearly the same
extrapolated value of $\alpha$. The data for ${\Delta H}_{max}=1$ are
more accurate and suggests $0.90\leq \alpha\leq 0.95$, which is consistent
with the previous analysis. The results for ${\Delta H}_{max}=2$ confirm the
trend to $\alpha<1$, although the uncertainties are larger.

Assuming the power-counting property (Eq. \ref{powercounting}) of the moments
of the width distribution (to be discussed in detail in Sec. V), we may also
use higher moments to estimate $\alpha$. The effective exponents obtained from 
$W_3$ have large fluctuations, but those obtained from $W_4$
behave similarly to the ones obtained from the interface width. They
are defined as
\begin{equation}
\alpha_{\left( L,i\right)}^{(4)} \equiv { \ln \left[W_{4,sat}\left( L\right)
/W_{4,sat}\left( L/i\right)\right] \over \ln{i} } ,
\label{defalphaL4}
\end{equation}
where $W_{4,sat}\left( L\right)$ are the fourth moments calculated at the
steady states.

\begin{figure}
\epsfxsize=8,5cm
\begin{center}
\leavevmode
\epsffile{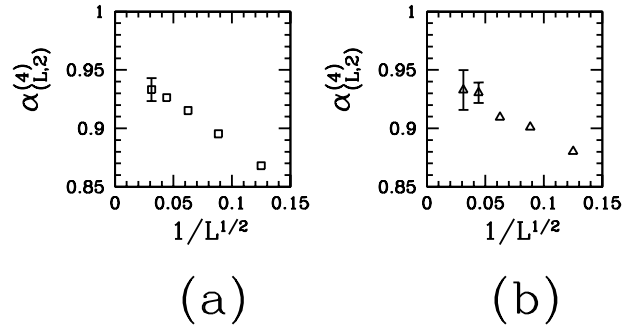}
\caption{Effective roughness exponents $\alpha_{\left( L,2\right) }^{(4)}$
(obtained from the fourth moment $W_4$) versus $1/L^{1/2}$ for
$1+1$-dimensional CRSOS models with (a) ${\Delta H}_{max}=1$ and (b) ${\Delta
H}_{max}=2$.  Error bars are shown only when they
are larger than the size of the data points.}
\label{fig4}                        
\end{center}
\end{figure}     

In Figs. 4a and 4b we show $\alpha_{\left( L,2\right)}^{(4)}$ versus
$1/L^{1/2}$ for the CRSOS models with ${\Delta H}_{max}=1$ and ${\Delta
H}_{max}=2$, respectively. The variable in the abscissa of Figs. 4a and 4b was
also chosen to illustrate the behavior of the data for large $L$ and not to
fit the data to a certain scaling form. The downward curvature of the plots
for large $L$ also suggest $\alpha<1$. The maximum and minimum
reasonable limits that can be inferred from the evolution of the data for
${\Delta H}_{max}=1$ give $0.92\leq\alpha\leq 0.96$. The accuracy of the
estimate for ${\Delta H}_{max}=2$ is lower, as before.

The intersection of at least two of the above estimates for
${\Delta H}_{max}=1$, obtained from the scaling of
different quantities and assuming different forms of finite-size corrections,
provides a final estimate $\alpha=0.94\pm 0.02$. As will be discussed below,
results for the DT model do not improve those obtained with the CRSOS model.

In Figs. 5a and 5b we show the effective exponents
$\alpha_{\left( L,2\right)}$ and $\alpha_{\left( L,2\right)}^{(4)}$ for the
noise-reduced DT model, also as a function of $1/L^{1/2}$. They are larger
than $\alpha=1$ and systematically increase with $L$. However,
from all previous theoretical work and the above numerical data for the CRSOS
models, there is no reason to expect $\alpha>1$ in the VLDS class.
Consequently, extrapolation of those data will not give reliable information
for the discussion on the exponents of the VLDS theory in $1+1$ dimensions. 
Instead, it is expected that the effective exponents for the noise-reduced DT
model (Figs. 5a and 5b) will eventually begin to decrease with $L$, possibly
for much larger $L$. Such decrease of $\alpha_{\left( L,2\right)}$ is actually
observed in the original DT model (without noise reduction), in the same range
of lattice lengths analyzed here~\cite{brunoc}. Also recall that, as shown in
Ref. \protect\cite{brunoc}, the data for original DT model also present huge
finite-size effects and cannot be used to obtain reliable estimates of VLDS
exponents.

\begin{figure}
\epsfxsize=8,5cm
\begin{center}
\leavevmode
\epsffile{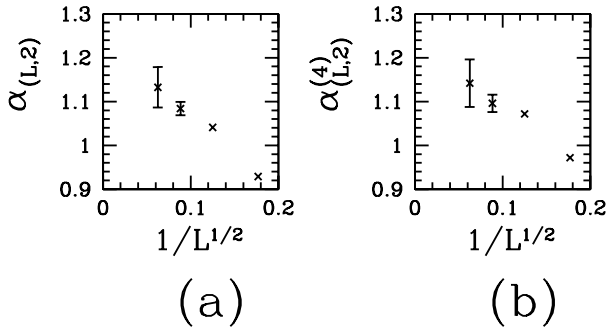}
\caption{Effective roughness exponents (a) $\alpha_{\left( L,2\right) }$
(obtained from the interface width) and (b) $\alpha_{\left( L,2\right)
}^{(4)}$ (obtained from $W_4$) versus $1/L^{1/2}$ for the $1+1$-dimensional DT
model. Error bars are shown only when they are larger than the size of the
data points.}
\label{fig5}                        
\end{center}
\end{figure}     

No improvement of the results in Figs. 5a and 5b is obtained by considering the
contribution of the intrinsic width (Eqs. \ref{intrinsic} and \ref{defalphai}).

There are other two points concerning our results for the DT model that
deserve some comments. The first one is the comparison with results of
Punyindu and Das Sarma in Ref. \protect\cite{punyindu}, who obtained
$\alpha\approx 1$ with noise reduction in lattice lengths $L\lesssim 60$.
Our effective exponents for the smallest lattices ($16\leq L\leq 64$)
correspond to two data points at the left sides (larger $1/L$) of Figs. 5a and
5b and those exponents are also near $\alpha=1$. Consequently, our estimates
are consistent with those of Ref. \protect\cite{punyindu}. On the
other hand, we conclude that the noise-reduction scheme works properly only in
a special range of lattice lengths, since its application to larger lattices
($L=128$ and $L=256$ in Figs. 5a and 5b) led to effective exponents larger
than $1$, indicating much more complicated finite-size behavior.

The other important point is related to the large error bars, particularly
for $L=256$. One of the reasons is certainly the relatively small number of
realizations for the largest lengths (see Sec. II). However, the
surfaces generated by the DT model in $d=1$ present grooves which may
survive during long times. These structures largely
increase the interface width of some realizations (see Ref.
\protect\cite{dasgupta1}) and, consequently, have remarkable influence on the
fluctuations of that quantity when averaged over various realizations.
However, note that this instability is controlled in the DT model, i. e. the
depths of the grooves do not diverge as time increases, contrary to other
discretized growth models which show true instabilities when pillars or
grooves are formed~\cite{dasgupta1,dasgupta2}. 

Now we turn to the calculation of the dynamical exponent.

Effective dynamical exponents are defined as
\begin{equation}
z_{\left( L,i\right)} = { \ln \left[ \tau_0\left( L\right) /\tau_0\left(
L/i\right)\right] \over \ln{i} } ,
\label{zL}
\end{equation}
so that $z_L\to z$ as $t\to\infty$. The error bars of $\tau_0$ are
larger than those of $\xi$ and the uncertainties are enlarged in the
calculation of effective exponents for small values of $i$ (Eq. \ref{zL}), then
we will work only with $i=4$.

\begin{figure}
\epsfxsize=8,5cm
\begin{center}
\leavevmode
\epsffile{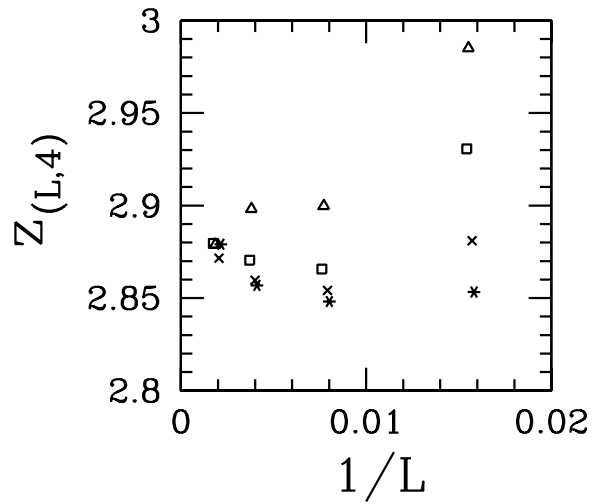}
\caption{Effective dynamical exponents $z_{\left( L,4\right) }$ versus $1/L$ for the
$1+1$-dimensional CRSOS model with ${\Delta H}_{max}=1$. Small horizontal
shifts of the data points were used to avoid their superposition. Error bars
(not shown) are smaller than $\Delta z=0.02$ (of this order for the largest
$L$).}
\label{fig6}                        
\end{center}
\end{figure}     

In Fig. 6 we show $z_{\left( L,4\right)}$ versus $1/L$ for the CRSOS model
with ${\Delta H}_{max}=1$, with $\tau_0$ calculated using four different
values of $k$ in Eq. (\ref{deftau0}) ($0.4\leq k\leq 0.7$). The data for
different $k$ clearly converge to the same region, providing an asymptotic
estimate $z=2.88\pm 0.04$. This final estimate also accounts for the error
bars (not shown in Fig. 6), which are near $\Delta z=0.02$ for the largest
values of $L$. Again it is clear that the value $z=3$ of one-loop
renormalization is excluded.

This conclusion is corroborated by the results for the CRSOS
model with ${\Delta H}_{max}=2$, although the accuracy of the data was poorer.
In Fig. 7 we show $z_{\left( L,4\right)}$ versus $1/L$ for that model, with
$\tau_0$ also calculated using four different values of $k$ in Eq.
(\ref{deftau0}).

Our results for the noise-reduced DT model do not provide useful information
on dynamical exponents, similarly to the case of the roughness exponents.

\begin{figure}
\epsfxsize=8,5cm
\begin{center}
\leavevmode
\epsffile{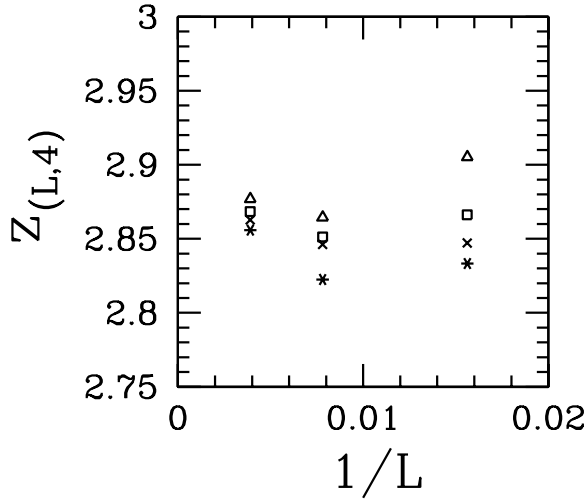}
\caption{Effective dynamical exponents $z_{\left( L,4\right) }$ versus $1/L$ for the
$1+1$-dimensional CRSOS model with ${\Delta H}_{max}=2$. Error bars
(not shown) are smaller than $\Delta z=0.03$ (of this order for the largest
$L$).}
\label{fig7}                        
\end{center}
\end{figure}     

\section{Scaling exponents in two-dimensional substrates}

In Figs. 8a and 8b we show $\alpha_{\left( L,2\right)}$ (Eq. \ref{defalphaL})
and $\alpha_{\left( L,2\right)}^{(4)}$ (Eq. \ref{defalphaL4}) for the
two-dimensional CRSOS model with ${\Delta H}_{max}=1$. Both linear fits give
$\alpha=0.662$, which is very near the one-loop renormalization value
$\alpha=2/3$ of the VLDS theory. Accounting for the error bars, which are
particularly large for $L=256$, we are not able to determine whether
$\alpha=2/3$ is exact or not. On the other hand, confirming other authors'
results~\cite{yook}, any difference from that value is
probably smaller than the two-loops correction of Janssen~\cite{janssen},
which is $\Delta\alpha\approx 0.014$.

Similarly to the one-dimensional case, the error bars
of the data for the model with ${\Delta H}_{max}=2$ are larger. Consequently,
no discrepancy from the one-loop exponents could be detected too.

The characteristic times $\tau_0$ for the model with ${\Delta
H}_{max}=1$ were obtained in lattices with $16\leq L\leq 128$, but their
values for the smallest lattices ($L=16$ and $L=32$) are very small, sometimes
below $\tau_0=1$ (one monolayer). For $L=256$, the accuracy of the interface
widths data is not enough to provide reliable estimates of $\tau_0$.
Consequently, we were not able to calculate accurate dynamical exponents in
the two-dimensional case. 

\begin{figure}
\epsfxsize=8,5cm
\begin{center}
\leavevmode
\epsffile{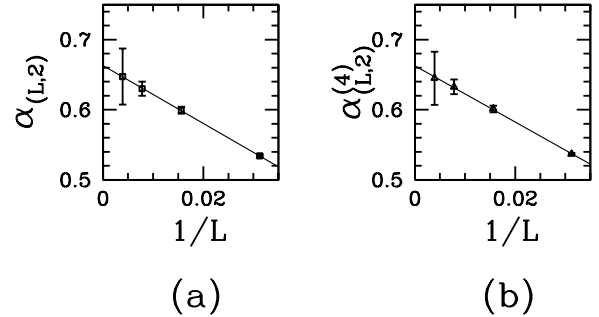}
\caption{Effective roughness exponents (a) $\alpha_{\left( L,2\right) }$
(obtained from the interface width) and (b) $\alpha_{\left( L,2\right)
}^{(4)}$ (obtained from $W_4$) versus $1/L$ for the $2+1$-dimensional CRSOS
model with ${\Delta H}_{max}=1$. Error bars are shown only when they are
larger than the size of the data points.}
\label{fig8}                        
\end{center}
\end{figure}     

\section{Universality of amplitude ratios}

Evidence on the power-counting property of the moments $W_n$ of the heights
distribution of VLDS models was given in Sec. III by the estimates of
$\alpha$ obtained from $W_2$ and $W_4$. A clearer evidence is given here by the
finite asymptotic estimates of the skewness and the kurtosis at the steady
states.

First we consider the models in $1+1$ dimensions.

In Figs. 9a and 9b we show the steady state skewness versus $1/L^{1/2}$ for
the CRSOS models with ${\Delta H}_{max}=1$ and ${\Delta
H}_{max}=2$, respectively. Except for the data for $L=1024$, which have
relatively large error bars, all points fall in
almost perfect straight lines, which give the asymptotic value $S=0.32\pm
0.02$ for both models.

In Figs. 9c and 9d we show the steady state kurtosis versus $1/L^{1/2}$ for
the CRSOS models with ${\Delta H}_{max}=1$ and ${\Delta H}_{max}=2$,
respectively. Only the data for $L\leq 512$ were shown because the
error bars are much larger for $L=1024$, not giving additional information on
the evolution of $Q$. Reasonable linear
fits are obtained with the last four data points in each case. The asymptotic
estimate is $Q=-0.11\pm 0.02$ for both models.

Our results for the competitive model (RD and CRSOS) introduced in Sec. II also
suggest that those amplitude ratios are universal for VLDS models. In that
case, there is no constraint on the difference of the heights of neighboring
columns, but only a trend to suppress large heights
differences. The coefficients $\nu_4$ and $\lambda_{4}$ in the
corresponding continuous equation (Eq. \ref{vlds}) are probably different from
those in the pure model ($p=0$), as obtained in related competitive
models~\cite{albano1,albano2}. In Figs. 10a and 10b we show, respectively,
$S\left( L,t\rightarrow\infty\right)$ and $Q\left(
L,t\rightarrow\infty\right)$ as a function of $1/L^{1/2}$ for the competitive
model. The asymptotic estimates are $S=0.32\pm 0.02$ and $Q\approx -0.1$,
which are near the previous estimates for the pure CRSOS model.

\begin{figure}
\epsfxsize=8,5cm
\begin{center}
\leavevmode
\epsffile{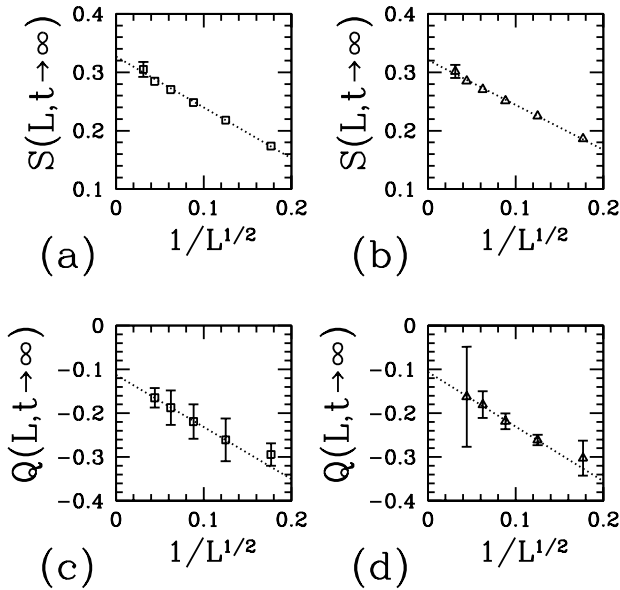}
\caption{Steady state skewness for the $1+1$-dimensional CRSOS model with
(a) ${\Delta H}_{max}=1$ and (b)${\Delta H}_{max}=2$, and steady state
kurtosis for that model with (c) ${\Delta H}_{max}=1$
and (d) ${\Delta H}_{max}=2$, as functions of $1/L^{1/2}$. Dotted lines are
least squares fits of the data. Error bars are shown only when they are
larger than the size of the data points.}
\label{fig9}                        
\end{center}
\end{figure}     

In Figs. 10c and 10d we show, respectively, $S\left(
L,t\rightarrow\infty\right)$ and $Q\left( L,t\rightarrow\infty\right)$ as a
function of $1/L^{1/2}$ for the noise-reduced DT model in $d=1$. There are
several reasons for the large error bars of the kurtosis, particularly in the
largest lattices. Firstly, as justified in Sec. III,
fluctuations in the data for the DT model are typically large. Secondly, the
relative fluctuations of the moments $W_n$ (Eq. \ref{defmoments}) rapidly
increase with the order $n$. Finally, while the size of the error bar of the
kurtosis is the same of $W_4/{W_2}^2$, the relative error significantly
increases when the constant $3$ is subtracted (Eq. \ref{defkurt}). The
relatively large errors in Figs. 9c and 9d (CRSOS models) can also be
explained along these lines.

\begin{figure}
\epsfxsize=8,5cm
\begin{center}
\leavevmode
\epsffile{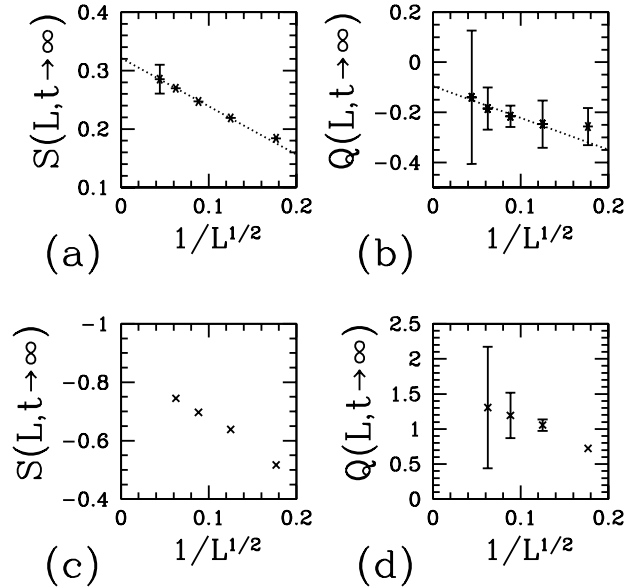}
\caption{(a), (b): steady state skewness and kurtosis, respectively, as a
function of $1/L^{1/2}$, for the competitive model (CRSOS with ${\Delta
H}_{max}=1$ and RD); (c), (d): steady state skewness and kurtosis,
respectively, as a function of $1/L^{1/2}$, for the DT model. Dotted lines are
least squares fits of the data. Error bars are shown only when they are
larger than the size of the data points.}
\label{fig10}                        
\end{center}
\end{figure}     

The trends of the data for the DT model in Figs. 10c and 10d are completely
different from those of the CRSOS models. We cannot exclude the possibility
that the universality of the amplitude ratios be a special feature of CRSOS
models and some simple extensions, like the above competitive model. However,
the behavior of the scaling exponents of the DT model is also unusual,
with no possible extrapolation to the expected region of the VLDS theory
($\alpha\leq 1$, $z\leq 3$), as discussed in Sec. III. Consequently, the
present results for the DT model, although not confirming the universality of
the amplitude ratios, are not reliable to discard that hypothesis (the negative
sign of the skewness is not a problem, since its sign changes with $\lambda_4$
- see related discussion in Ref. \protect\cite{kpz2d}).

Now we turn to the CRSOS models in $2+1$ dimensions.

In Figs. 11a and 11b we show the steady state skewness versus $1/L^{1/2}$ for
the CRSOS models with ${\Delta H}_{max}=1$ and ${\Delta
H}_{max}=2$, respectively. The asymptotic estimates are $S=0.19\pm 0.02$ and
$S=0.20\pm 0.02$, which also suggest the universality of this quantity.
In Figs. 11c and 11d we show the steady state kurtosis versus $1/L^{1/2}$ for
the CRSOS models with ${\Delta H}_{max}=1$ and ${\Delta H}_{max}=2$,
respectively. The asymptotic value $Q=0$, which is the Gaussian value, is
consistent with the error bars. Thus, in $2+1$ dimensions, we also obtain
evidence of universality of the amplitude ratios for CRSOS models, which
suggests this possibility for the whole VLDS class.

\begin{figure}
\epsfxsize=8,5cm
\begin{center}
\leavevmode
\epsffile{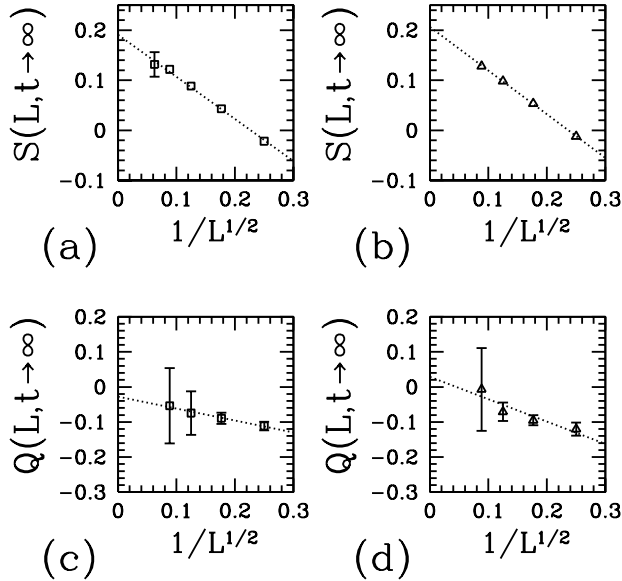}
\caption{Steady state skewness for the $2+1$-dimensional CRSOS model with
(a) ${\Delta H}_{max}=1$ and (b)${\Delta H}_{max}=2$, and steady state
kurtosis for that model with (c) ${\Delta H}_{max}=1$
and (d) ${\Delta H}_{max}=2$, as functions of $1/L^{1/2}$. Dotted lines are
least squares fits of the data. Error bars are shown only when they are
larger than the size of the data points.}
\label{fig11}                        
\end{center}
\end{figure}     

\section{Summary and conclusion}

We studied numerically discrete growth models which belong to the VLDS class
in $1+1$ and $2+1$ dimensions. Scaling exponents and steady state values of the
skewness and the kurtosis, which characterize the heights
distribution, were determined for those models.

Results for the CRSOS model with ${\Delta H}_{max}=1$ gave the
roughness exponent $\alpha=0.94\pm 0.02$ and the dynamical exponent $z=2.88\pm
0.04$ in $d=1$. These estimates confirm the proposal of Janssen~\cite{janssen}
that the exponents of the VLDS theory obtained from one-loop renormalization
($\alpha=1$ and $z=3$) are not exact. The corrections from two-loops
calculations give $\alpha\approx0.97$ and $z=2.94$, but they are obtained
from expansions in $4-d$, which are not expected to provide accurate results
for small $d$. On the other hand, the negative sign of the correction to
one-loop results is consistent with our findings. In $d=2$, our
results are not able to exclude the one-loop values, confirming
other authors' conclusions~\cite{yook}.

The estimates of the steady state skewness and kurtosis of the CRSOS models
with ${\Delta H}_{max}=1$ and ${\Delta H}_{max}=2$ and of the competitive
model (RD versus CRSOS with ${\Delta H}_{max}=1$) suggest that those
amplitude ratios are universal in the VLDS class. However, for the DT model
in $d=1$, which belongs to the same class, those quantities are very
different from the suggested universal values. One possible reason for this
discrepancy is the slow convergence of the DT data to the VLDS behavior. The
hypothesis of a slow crossover is supported by the fact that the estimates of
$\alpha$ for the DT model are significantly larger than the values predicted
theoretically and confirmed numerically ($\alpha\leq 1$ in $d=1$). Another
possibility is that both CRSOS models and the
competitive model have continuum representations with suitable combinations of
coefficients which lead to the same forms of the heights distributions.

We believe that the results of this work will motivate further studies,
numerical and analytical, of the VLDS equation and related discrete models. The
estimates of scaling exponents in $d=1$ and the apparent universality of
amplitude ratios are some of the results that may eventually help one to
validate approximations in analytical works. On the other hand, numerical
solutions of the VLDS equation or simulations of new discrete models in this
class would be relevant to broaden the present discussion. 

\vskip 1cm

{\bf Acknowledgements}

The author thanks useful suggestions of Prof. H. K. Janssen and Prof. S. Das
Sarma.

This work was partially supported by CNPq and FAPERJ (Brazilian agencies).

\end{document}